\documentstyle[twoside,fleqn,espcrc2,epsfig]{article}

\newcommand{\co}{ ,}
\newcommand{\nn}{\nonumber \\}
\newcommand{\scs}{\co \;}
\newcommand{\per}{ \; .}

\newcommand{\sem}{\;\; ; \; }

\newcommand{\calleff}{{\cal L}_{\mbox{\small{eff}}}}

\newcommand{\boldphi}{{\mbox {\boldmath $\pi$}}}

\newcommand{\ed}{\end{document}}
\newcommand{\be}{\begin{equation}}
\newcommand{\ee}{\end{equation}}
\newcommand{\bea}{\begin{eqnarray}}
\newcommand{\eea}{\end{eqnarray}}
\newcommand{\fs}{\per}
\newcommand{\la}{\langle}
\newcommand{\ra}{\rangle}
\title{\begin{flushright}{\small{BUTP-99/32}\\December 1999}\end{flushright}
Chiral perturbation theory
\thanks{Invited talk given at the QCD Euroconference 99, Montpellier,
 7-13th July,
1999. The present article  contains additional  references as
compared to the version that will appear in the proceedings.}
\thanks{
This work  was supported in part by the Swiss National Science
Foundation, and by TMR, BBW--Contract No. 97.0131  and  EC--Contract
No. ERBFMRX--CT980169 (EURODA$\Phi$NE).}\\[.5cm]}

\author{J. Gasser\address{
Institut f\"ur  Theoretische Physik, Universit\"at  Bern,
Sidlerstrasse 5, CH--3012 Bern, Schweiz \\
\hspace{.3cm}e-mail: gasser@itp.unibe.ch}}

\begin{document}

\begin{abstract}
\begin{center} { Abstract}
\end{center}
I present an outline of chiral perturbation theory and discuss
  some recent developments in the field.\\[.5cm]
{\bf Pacs}: 11.30.Rd, 12.39.Fe
\end{abstract}

\maketitle

\section{EFFECTIVE THEORY}
 The QCD
lagrangian can be replaced at low energy with
an effective lagrangian that is formulated  in terms of the
asymptotically
observable fields \cite{weinberg,glan,hlan}.
 This effective lagrangian reads for
processes with pions alone
 \bea\label{effmm}
{\cal L}_M =\frac{F^2}{4}\langle \partial_\mu U \partial ^\mu U^\dagger +
M^2(U+U^\dagger)\rangle\per
\eea
Here, the matrix field $U$ is an element of $SU(2)$,
and the symbol $\langle A \rangle$ denotes the trace of the
matrix $A$.
 In the
following, I use  the
parameterization
 \bea
U&=&\sigma +\frac{i\phi}{F}\sem\phi=\left(\begin{array}{cc}
\pi^0&\sqrt{2}\pi^+\\
\sqrt{2}\pi^-&-\pi^0\end{array}\right)\co \nonumber\\
\sigma&=&\left[{\mbox{\bf 1}}-\phi^2/F^2\right]^{\frac{1}{2}}\co
\eea
and the notation
\bea
\phi&=& \sum_{i=1}^3\tau^i\phi_i\scs
\boldphi=(\phi_1,\phi_2,\phi_3)\per
\eea
The coupling constant $F\simeq 93$ MeV measures the strength of the $\pi\pi$
interaction, and the quantity $M^2$ denotes the square of the
  physical pion mass
(that I denote with $M_\pi$) at lowest order in an expansion in powers of
$1/F$, see below.
It is proportional to the light quark
masses $m_u,m_d$,
\bea
M^2=2\hat{m}B\scs \hat{m}=\frac{1}{2}(m_u+m_d)\co
\eea
where $B$ itself is related to the quark condensate, see \cite{glan}.
Note that the quantity $M^2$ occurs not only in the kinetic term
of the pion lagrangian, but also in the interaction: it acts
both as a mass parameter and as a coupling constant.
 The
lagrangian
${\cal L}_M$ is called the "non-linear sigma-model lagrangian".
This name has led to some confusion in the
literature about the
meaning of the effective lagrangian:
one is not replacing QCD
with a "chiral model", as this procedure is often called.
To the contrary, ${\cal L}_M$ can be used to
calculate processes at low energies, with a result that is
identical  to the one
in QCD \cite{weinberg,glan,hlan}.

\begin{figure}[h]
\begin{center}
\mbox{
       \epsfig{file=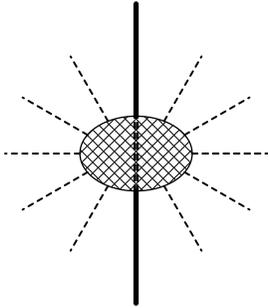,height=4cm}
     }
     \end{center}
 \caption{
The nucleon traveling through space, emitting and absorbing pions.
}
\label{fig1}
\end{figure}

In case we wish to consider also nucleons, one has to enlarge the above
lagrangian.
Let us consider processes where a single baryon
(proton or neutron) travels in space, emitting and absorbing pions
in all possible ways allowed by chiral symmetry. This process is
illustrated in  Fig. \ref{fig1}.
One need not consider processes with closed nucleon lines.
These contributions may be absorbed in a renormalization of the
coupling constants in the effective lagrangian for meson-nucleon interactions,
\bea
{\cal L}_{MB}&=&\bar{\Psi}
\left\{i\not\hspace{-1.1mm}D-m+\frac{1}{2}g_A
\not \hspace{-.82mm}u  \gamma_5\right\}
\Psi\co\label{effmn}
\eea
with
\bea
u^2=U, u_\mu=iu^\dagger\partial_\mu Uu^\dagger,\nonumber\\
\Gamma_\mu=\frac{1}{2}[u^\dagger,\partial_\mu u],
D_\mu=\partial_\mu+\Gamma_\mu\per
\eea
Here, $\Psi$ denotes the nucleon field, $m$ is the nucleon mass in
the chiral limit $m_u=m_d=0$, and $g_A$ is the neutron
 decay constant $g_A\simeq 1.25$.
The
effective lagrangians (\ref{effmm}),(\ref{effmn}) contain the three
couplings $1/F,M^2$, $g_A$ and the nucleon mass $m$ as free parameters.
These couplings are not fixed by chiral symmetry,
\bea
F=c_1\Lambda_{QCD}\scs m=c_2\Lambda_{QCD}\ldots\co
\eea
where $\Lambda_{QCD}$ is the renormalization group invariant scale of
QCD, and where $c_i$ are dimensionless numbers that can in principle
be calculated in QCD. On the other hand, there are {\em relations}
between {\em physical} quantities, e.g., the famous Goldberger Treiman
relation
\bea
g_{\pi N}=\frac{m g_A}{F}\per
\eea
The quantities $g_{\pi N}, F, m$ and $g_A$ are evaluated in the chiral
limit $m_u=m_d=0$. In the real world, there are corrections of order
$m_u,m_d$ to this relation \cite{goldberger}.

\section{TREE GRAPHS}
Tree graphs evaluated with (\ref{effmm}),(\ref{effmn})
generate the leading order term in a systematic low-energy expansion of
the $S$ - matrix elements
\cite{weinberg,glan,hlan}.
 I illustrate this fact  with two examples.

\subsection{The pion mass}
It suffices to consider the terms
in ${\cal L}_M$ that are quadratic in the pion fields,
\bea
{\cal L}_M =\frac{1}{2}\left[\partial_\mu \boldphi \cdot \partial^\mu \boldphi
-M^2\boldphi^2\right] +\cdots \per
\eea
Therefore, the effective theory contains at tree level three
mass degene\-rate bo\-sons $\pi^+,\pi^-,\pi^0$, with
\bea
M_{\pi^+}^2=M_{\pi^-}^2=M_{\pi^0}^2= M^2\per\label{eqpionmass2}
\eea
At the leading order considered here, there is no isospin
splitting:
the masses of the charged and of the neutral pion are identical, see
 \cite{weinberg}.
A small mass
difference due to $m_u\neq m_d$ does show up only
 at next order in the chiral expansion.

\subsection{$\pi\pi$ scattering}
The full power of the effective lagrangian method comes into play when one
starts to evaluate scattering matrix elements. Consider for this purpose
elastic $\pi\pi$ scattering. The interaction part of the effective lagrangian
is
 \bea
{\cal
L}_{int}&=&\frac{1}{8F^2}\left\{\partial_\mu\boldphi^2\partial^\mu
\boldphi^2 -M^2(\boldphi\cdot\boldphi)^2\right\}\nn
&&+\cdots\per \eea
Since we calculate tree matrix elements, the terms at order
O($\boldphi^6$) - indicated by the ellipses - are not needed. The
contributions with four
fields in the lagrangian contain two types of vertices: the first one has two
derivatives, while the second contains the parameter $M^2$ as a coupling
constant. In the following I consider the isospin symmetry limit $m_u=m_d$
and use the standard notation
\bea
T^{ab;cd}=\delta^{ab;cd}A(s,t,u) + {\mbox{cycl.}}
\eea
for the matrix element of the process
\bea
\pi^a(p_1)\pi^b(p_2)\rightarrow \pi^c(p_3)\pi^d(p_4)\co
\eea
with the  Mandelstam variables
\bea
s=(p_1+p_2)^2\co t=(p_1-p_3)^2\co\nn u=(p_1-p_4)^2\per
\eea
 The result of the calculation is
\bea
A
\stackrel{\mbox{\small{tree}}}{=}\frac{s-M^2}{F^2}
\stackrel{\mbox{\small{tree}}}{=}
\frac{s-M_\pi^2}{F_\pi^2}
\per\label{eqatree}
\eea
The second equal sign in Eq. (\ref{eqatree}) is based on the fact
 that the coupling $M^2$ can be replaced at tree level with
the square of the physical pion mass, see Eq. (\ref{eqpionmass2}),
and that the physical pion decay constant $F_\pi$ is equal to $F$
 in the same approximation. Of course, the result Eq. (\ref{eqatree}) agrees
 with the expression evaluated \cite{weinpipi} with current algebra techniques
 a long time ago.

In order to compare the above expression for the scattering matrix element with
the data, it is useful to consider the partial wave expansion of the amplitude.
 I consider the isospin zero combination
\bea
&&\hspace{-.6cm}T^0(s,t) =3A(s,t,u)+A(t,u,s)+A(u,s,t)\co\nn
&&\hspace{-.6cm}s=4(M_\pi^2+p^2)\co
 t=-2p^2(1-\cos \;\theta\;)\co
\eea
where $\theta$ is the scattering angle in the center of mass system, and $p^2$
is the square of the pion momentum.  $T^0(s,t)$
may  be expanded in Legendre polynomials,
\bea\label{eq:partial}
T^0(s,\theta)=32\pi\sum_{l=0}^\infty(2l+1)P_l(\cos\;\theta)
t_l^0(s)\co
 \eea
with energy-dependent coefficients $t_l^0(s)$. Unitarity implies
that, in the elastic region
$4M_\pi^2 \leq s \leq 16M_\pi^2$, the
coefficients have  the structure
\bea
&&t_l^0(s)\stackrel{\mbox{\small{unitarity}}}{=}\frac{1}{\sigma}
e^{i\delta_l^0(s)} \sin\delta_l^0(s)\co\nn
&&\hspace{1cm}\sigma=(1 - 4M_\pi^2/s)^{1/2}\co
\label{eqdelta}
\eea
with real phase shifts $\delta_l^0$. Therefore, knowing
$A(s,t,u)$, one may evaluate $T^0$, then $t_l^0$ and finally the
phase shift $\delta_l^0(s)$ in the low-energy expansion.
 The behavior of the partial wave amplitudes
near threshold is of the form
\bea
{\Re}e \; t_l^I(s)=p^{2l}\{a_l^I+p^2 b_l^I +O(p^4)\}\per
\eea
The tree-level result (\ref{eqatree}) gives~\cite{weinpipi}
\bea
a_0^0\stackrel{\mbox{\small{tree}}}{=}\frac{7M_\pi^2}{32\pi
F_\pi^2 }=0.16\co
\eea
to be compared with the observed value~\cite{rosselet,frogg}
 \bea
a_0^0 \stackrel{{\mbox{\small{exp.}}}}{=}0.26 \pm
0.05\per \eea

\section{LOOPS}
The isospin
zero amplitude in elastic $\pi\pi$ scattering is
real at tree level,
 \be
T^0\stackrel{\mbox{\small{{tree}}}}{=}\frac{2s-M_\pi^2}{F_\pi^2}
\per\label{eqa0tree}
\ee
On the other hand, unitarity (\ref{eqdelta}) requires the
Legendre coefficients in the
partial wave expansion (\ref{eq:partial}) to be complex.
 This apparent inconsistency
arises  for the following reason.
CHPT represents the amplitude $T^0$ through an energy expansion - the
tree-level result (\ref{eqa0tree}) is the leading order term, quadratic in
the momenta. The
 partial waves  $t_l^0$ are therefore  also of order $p^2$
according to Eq. (\ref{eq:partial}).
One concludes furthermore from
\bea
{\Re }e\; t_l^0=\frac{1}{2\sigma}\sin{2\delta_l^0}
\eea
that the phase shifts
$\delta_l^0$ are of order $p^2$ as well. The imaginary part of
the partial waves,
 \bea
{\Im }m \;t_l^0=\frac{1}{\sigma(s)}\sin^2{\delta_l^0}\co
\eea
is then of  order $p^4$,
 as is the imaginary part of the amplitude $T^0$.
 Since we have not yet
considered the amplitude to this accuracy,  we have simply missed
its absorptive  part so far.

The remedy is simple: one  needs to consider loops, generated by
the effective lagrangian (\ref{effmm}) \cite{weinberg}.
The amplitudes so evaluated satisfy
unitarity in a perturbative sense. Moreover, they have the correct analytic
and crossing symmetry properties.

The expansion according to the  number of
independent loops
of connected Feynman diagrams may be identified with an expansion in inverse
powers of $F^2$. Indeed, by definition, the number of independent loops
generated by ${\cal L}_M$ is the number
of independent four-momenta in the diagrams.
 The effective lagrangian (\ref{effmm})
has an infinite number of vertices. In the following I denote by
$V_n$ the
number of vertices with $n$ fields in the diagram under consideration, and by
$I$ the number of internal lines. Using energy-momentum
conservation at each
vertex, the number $L$ of independent loops is
\bea
L=I+1-\sum_n V_n\fs
\label{loop2}
\eea
Furthermore, for connected {\it tree} graphs the number $E$ of
external lines is \bea
E=\sum_n nV_n-2I\fs
\label{loop3}
\eea
This formula is valid also for connected diagrams with loops.
Finally, eliminating $I$ in
 Eq. (\ref{loop3}) with the help of Eq. (\ref{loop2}) gives
\bea
E+2L=2 +\sum_n (n-2)V_n\fs\label{eqe2l}
\eea
Next, I count powers of $F^2$.
 Each
vertex that contains $n$ fields generates a factor
$F^{2-n}$. As a result, the overall power of $F$ in a fixed
diagram is $\sum_n(2-n)V_n$. According to Eq. (\ref{eqe2l}),
the diagram is therefore proportional to
\bea
\frac{F^{2-E}}{F^{2L}}\fs\label{eqpowerf}
\eea
This shows that  the loop expansion coincides with the expansion in
inverse powers of $F^2$. For a connected $n$-point function
$G_n$, the expansion reads
\bea
&&\hspace{-.5cm}F^{n-2}G_n=\nn
&&\hspace{-.5cm}G_{n,{\mbox{\small{tree}}}}
+\frac{G_{n,{\mbox{\small{1 loop}}}}}{F^2}
 +\frac{G_{n,{\mbox{\small{2 loops}}}}}{F^4}
+\cdots \fs
\eea
For dimensional reasons, $n$ loop contributions are
therefore suppressed by
$2n$ powers of energy with respect to the tree diagram. In
dimensional regularization, the only dimensionful
parameters in the effective theory - besides $F$ - are the external momenta
and $M^2$. We conclude
that the loop expansion amounts to an expansion in powers of the external
momenta and of $M^2$, where each term in this expansion is
multiplied with a dimensionless function of the momenta and of
$M^2$. Therefore, the loop expansion is equivalent to an energy
expansion\cite{weinberg}.

\section{EXTERNAL FIELDS}
For the evaluation of loops, it is useful to introduce the concept of
 external fields.
 Let us consider QCD in
the two
flavor case, and define
\bea
{\cal L}&=&{\cal L}_{QCD}^0+\triangle {\cal L}\co\nn
\triangle{\cal L}&=&
\bar{q}\gamma_\mu\left[v^\mu(x)+\gamma_5a^\mu(x)\right]q \nn
&&-\bar{q}\left[s(x) - i\gamma_5 p(x)\right]q\per \label{eqgenqcd}
\eea
The symbol ${\cal L}_{QCD}^0$ denotes the QCD lagrangian without the quark mass
matrix. The external fields $v_\mu,a_\mu,s$
and
$p$  are hermitian, color neutral  two by two matrices in
flavor space. In order to avoid the discussion of
anomalies~\cite{wesszumino}, I consider in the following
only the case where the external vector and axial fields are
traceless,
\bea
\langle v_\mu\rangle &=& \langle a_\mu\rangle = 0 \per
\eea
The generating functional $\Gamma$ is given by \bea
e^{i\Gamma(v,a,s,p)}=\langle 0|Te^{i\int d^4\!x\triangle {\cal
L}}|0\rangle\per
\eea
It contains all the information on the
Green functions
built from vector, axial, scalar and pseudoscalar quark currents.
As an example, the term linear in the scalar field,
\bea
\Gamma=-\int d^4\!xs^{\alpha \beta}\langle
0|\bar{q}^\alpha(x)q^\beta(x)|0\rangle
+\cdots\co\label{eqcondensate} \eea
contains the vacuum expectation value of the quark fields, whereas the term
quartic in the axial current $a^\mu$ contains the $\pi\pi$ scattering
amplitude, and so on. By expanding $\Gamma$ around
$v^\mu=a^\mu=s=p=0$ one generates the
Green functions in the chiral limit $m_u=m_d=0$, whereas the expansion around
\bea
v^\mu=a^\mu=p=0, s={\mbox{diag}}(m_u,m_d)\label{eqexp}
\eea
generates the Green functions at finite values of the quark
masses.

The generating functional $\Gamma$ contains the complete knowledge of
Green function built from quark currents - it is therefore impossible
to evaluate it in closed form with present techniques.
On the other hand,
the invariance theorem
proven  by
Leutwyler \cite{hlan}   states that $\Gamma$
may be evaluated at low energies using an effective theory, where only the
observed asymptotic states occur in the lagrangian. In addition, the effective
lagrangian may be taken to be  gauge invariant by itself. The
corresponding rules to evaluate the Green functions of quark
currents in QCD with two flavors are discussed in the following
section.

\section{EFFECTIVE THEORY OF QCD}

The central object in the invariance theorem \cite{hlan} is the gauge
invariant effective
lagrangian. It consists of a series of terms, each of which is gauge invariant
by itself,
\bea
{\cal L}_{\mbox{\small{eff}}}={\cal L}_2 +{\cal L}_4 + {\cal L}_6
+\cdots \; \per
\eea
Here, ${\cal L}_{2n}$ contains $m_1$  derivatives and
$m_2$ quark mass
matrices, with $m_1+2m_2=2n$ (I  consider here standard power counting for
the chiral condensate - the generalized case is discussed below).
The leading
term in the low-energy expansion is obtained by evaluating tree graphs with
${\cal L}_2$. The next-to-leading contributions are obtained by evaluating
one-loop graphs with ${\cal L}_2$ and tree graphs generated by
${\cal L}_2+{\cal L}_4$ with exactly one vertex from ${\cal L}_4$,
etc. This procedure to evaluate Green functions is called chiral
perturbation theory (CHPT).

In order to construct these effective lagrangians, it is useful to first have
building blocks that transform covariantly under local
gauge
transformations. For this purpose,
one  defines the transformation $h(x)$ by
\bea
U&&\stackrel{G}{\rightarrow} V_R U V_L^\dagger \co\nn
u&&\rightarrow V_R u h^\dagger\, , \; u^2=U\per
\eea
Furthermore, one uses the field
         \bea
         \chi = 2 B(s+ i p)\co
         \eea
 the covariant derivative
\bea
  D_\mu X= \partial_\mu X- i (v_\mu+a_\mu) X + i X (v_\mu-a_\mu)
,
 \eea
and the   building blocks
\bea
         u_\mu &=& i u^\dagger D_\mu Uu^\dagger = -i u D_\mu U^\dagger u =
u_\mu^\dagger \co
         \nn
         \chi_\pm &=& u^\dagger \chi u^\dagger \pm u \chi^\dagger u \co \nn
         \chi_-^\mu&=& u^\dagger D^\mu\chi u^\dagger -uD^\mu\chi^\dagger u \co
\nn
         f^{\mu \nu}_\pm &=& u F^{\mu \nu}_L u^\dagger \pm u^\dagger F^{\mu
\nu}_R
         u \per
          \eea
 The quantity $F^{\mu \nu}_R \; (F_L^{\mu \nu})$ stands
 for the field strength associated with the nonabelian external field
         $v_\mu + a_\mu \; (v_\mu - a_\mu)$.
 Each of the above building blocks
transforms as \bea
         I \stackrel{G}{\rightarrow} hIh^\dagger
\eea
under {\it local} gauge transformations.

The effective lagrangians ${\cal
L}_{2,4}$ are \cite{glan}
 \bea
         {\cal L}_2 &=& \frac{F^2}{4} \la u_\mu u^\mu
 + \chi_+\ra\co
\eea
and
\bea
      {\cal L}_4 = \sum^{7}_{i=1} l_iP_i + \cdots
\co\label{eql4} \eea
where
\bea
&&\hspace{-1cm}\begin{array}{ll}
&\\
         P_1 = \frac{1}{4} \la u^\mu u_\mu \ra^2 \co &
         P_2 = \frac{1}{4} \la u_\mu
         u_\nu\ra \la u^\mu u^\nu \ra \co
         \\
&\\
         P_3 = \frac{1}{16} \la \chi_+\ra^2 \co &
         P_4 = \frac{i}{4} \la u_\mu
         \chi^\mu _- \ra \co \\
&\\
         P_5 = - \frac{1}{2} \la f_-^{\mu \nu} f_{- \mu \nu} \ra \co &
         P_6 =
         \frac{i}{4} \la f^{\mu \nu}_+ [u_\mu, u_\nu]\ra \co
         \\
&\\
         P_7 = -\frac{1}{16}\la \chi_- \ra^2\per&\\
\end{array}          \label{eql41}
\eea
For $a_\mu=v_\mu=p=0, s=$ diag($m_u,m_d$), the lagrangian ${\cal L}_2$
agrees with ${\cal L}_M$ in (\ref{effmm}).
The ellipses in (\ref{eql4}) denote polynomials in the external fields
which are independent of the pion variables. These do not contribute to
$S$-matrix elements.
The lagrangian contains 7 low-energy constants $l_i$ (LEC's).
Some of these are
divergent in four dimensions - they cancel the divergences generated
by the one-loop graphs \cite{glan}. The structure of ${\cal L}_6$ will
be discussed below.

\section{APPLICATIONS}
\subsection{Theory}
By use of the above technique, most matrix elements accessible to experimental
data have been evaluated to one loop accuracy, including baryons and
weak interactions. It is  impossible to cover here the results
of  these investigations. I refer  the interested reader instead to
the {\em Second DA$\Phi$NE Physics Handbook} \cite{dafnesecond} and to
recent chiral workshops \cite{mainz,honnef} for a collection of
many results. In addition, let me mention the EURODA$\Phi$NE
collaboration \cite{eurodafne}, where 10 European universities
and research institutes
have formed a network, in order to study  high precision elementary
particle physics at the DAFNE $\Phi$-factory. Topics considered by the
network include
\begin{itemize}
\item
CP and CPT physics in the kaon system
\item
Chiral Perturbation Theory
\item
$K$ and $\eta',\eta$ decays
\item
$K_{l3},K_{l4}$ decays, $\pi\pi\rightarrow \pi\pi$
\item
total hadronic cross-section in electron-positron collisions below 2
GeV
\item
vector mesons in effective lagrangians
\item
nuclear physics with kaons from $\Phi$ decays
\end{itemize}

I refer the interested reader to the relevant Home Pages
\cite{eurodafne} for more information.

\subsection{Experiment}
There are presently  several experimental activities involved
in the low-energy region of the Standard Model. The following table
lists some of them.

\vskip.5cm

\hspace{-.55cm}\begin{tabular}{ll}
DIRAC &\hspace{.4cm}$\pi^+\pi^-$ - atom \\
(CERN)&\hspace{.4cm}{\footnotesize{$\pi\pi$ scattering lengths}}\\
&\\
KLOE &\hspace{.2cm}
$K_{l3},K_{l4},K\rightarrow \pi\pi, \eta\rightarrow 3\pi,...$\\
(Frascati)&\hspace{.4cm}{\footnotesize{$\pi\pi$ scattering
lengths from $K_{l4}$}}\\
&\\
DEAR  &\hspace{.4cm}$KN$ - atom \\
(Frascati)&\hspace{.4cm}{\footnotesize{$K N$ scattering lengths}}\\
&\\
MAMI &\hspace{.4cm}$\gamma N\rightarrow \pi N,...$\\
(Mainz)&\\
&\\
E865 &\hspace{.4cm}
$ K_{l3},K_{l4},...$\\
(Brookhaven)&\hspace{.4cm}{\footnotesize{$\pi\pi$ scattering lengths from
$K_{l4}$}}\\
&\\
R-98-01.1 &\hspace{.4cm}$\pi p$ - atom\\
(PSI)&\hspace{.4cm}{\footnotesize{$\pi N$ scattering lengths}}\\
\end{tabular}

\vskip.5cm

We expect that these experiments will provide us with relevant new
insight into the low-energy structure of the Standard Model.

\section{SOME RECENT DEVELOPMENTS}
At this conference, there were several talks related to recent
developments in low-energy effective theories: E. de Rafael (large
$N_c$), L. Girlanda (chiral condensate), M. Knecht (large $N_c$),
S. Peris (large $N_c$), J. Prades ($\triangle I$ = 1/2 rule),
H. Sazdjian (pionium), J. Soto (pionium). I refer the reader to the
corresponding contributions in these proceedings for detailed
information. In the following, I present several topics where
progress has recently been achieved - of course, this selection
 is a matter of personal taste.

\subsection{Effective lagrangian at $O(p^6)$}
Chiral perturbation theory in the meson sector is now being carried
out at next-to-next-to-leading order. Several complete two-loop
calculations exist \cite{p6su2,p6su3}.
To relate the low-energy constants that occur at order
$p^6$ to those appearing in other
processes, one needs to know the effective
lagrangian ${\cal L}_6$ in its most general form. It has been
constructed recently for the general flavor case, as well as for
$N_f=2,3$ \cite{lp6fs,lp6bce}. In the case of $N_f$  light flavors,
there are 112 in principle measurable and 3 contact terms, that
 reduce to 90+4 (53+4) for 3 (2) flavors \cite{lp6bce}. In
Ref. \cite{lp6bce}, the divergence structure of ${\cal L}_6$ has been
determined as well. This provides a very thorough check on any
specific  two-loop calculation.

The number of new couplings may seem large. On the other hand, in
the chiral limit $m_u=m_d=0$, the number on new phenomenological
constants goes down \cite{lp6bce}:
\bea
\hspace{.5cm}\begin{tabular}{lrl}
3&{\mbox{LEC's in }}&$\pi\pi\rightarrow\pi\pi$\\
6&{\mbox{ in }}&$\gamma\gamma\rightarrow\pi\pi$\\
3&{\mbox{ in }}&$\tau\rightarrow 3\pi\nu_\tau$\\
2&{\mbox{in }}&$\pi\rightarrow l\nu\gamma$\\
2&{\mbox{in }}&$F_V^\pi(t)$\\
1&{\mbox{in }}&$\pi\rightarrow l\nu\gamma^*\per$\\
\end{tabular}
\eea
This is quite a manageable number of terms. Still, it remains to
be seen whether they allow one to relate different observables in a
useful and practical manner. In addition, one may rely on  the resonance
exchange approximation to estimate some of the relevant constants at
this order \cite{resonance,p6su2}, or on  sum rules
\cite{sumrule}.

\subsection{Radiative corrections}
Once experimental data are sensitive to two-loop contributions, one is
forced to also consider radiative corrections due to virtual photons.
 It has been shown \cite{radstrongpion} that the electromagnetic corrections
to the S-wave scattering lengths are of comparable size to the
$O(p^6)$ strong interaction contributions.
  The relevant effective
lagrangian in the strong interaction sector - including real and
virtual photons - has been investigated, and several calculations
have already been
performed \cite{radstrongpion,radstrongnucleon}. In addition, for the
not so rare kaon
decays being investigated by E865 \cite{e865} and by KLOE \cite{kloe},
it will be important to
be able to perform these corrections in a systematic manner also
for  weak interactions. The
 relevant effective lagrangian has recently been constructed \cite{radweak}.

\subsection{Baryon CHPT}
Chiral perturbation theory in the
baryon sector is not as straightforward as in the meson sector,
because the baryon mass does not vanish in the chiral limit and
generates thus a new scale. Power counting becomes more difficult -
 the contribution from loops is not automatically suppressed at low
energy \cite{gss}.
In the last decade, a special method has been set forward - referred
to  as heavy baryon chiral perturbation theory (HBCHPT)- to cope with this
problem \cite{hbchpt}. Recently, it has been shown
that
 one may  stay in a manifestly Lorentz invariant framework
 by simply treating the Feynman integrals in an appropriate  manner,
 such that the
infrared singular pieces are singled out automatically, and the
polynomial terms that set up the power counting are discarded
 \cite{tang,becherl}. Using
this method, it has
been shown \cite{tang} that the one-loop expression
for pion nucleon scattering, worked out  a long time ago \cite{gss} in
the framework of relativistic chiral perturbation theory,  has a
low-energy expansions at order $p^3$ that is identical to the
 one performed in HBCHPT (in the kinematic region  where HBCHPT is
applicable \cite{becherl}).

\subsection{Elastic $\pi\pi$ scattering and Roy equations}

The interplay between theoretical  and experimental
aspects of elastic $\pi\pi$ scattering
is illustrated in figure 2.
 As we discussed in subsection 2.1, Weinberg's calculation \cite{weinpipi}
of the scattering amplitude  at leading order in the
low-energy expansion gives for the isospin zero S-wave scattering
 length the value
$a_0^0=0.16$ in units of the charged pion
mass. This  differs from the experimentally determined
value \cite{rosselet}
$a_0^0=0.26 \pm 0.05$ by two standard deviations. The one-loop
calculation \cite{glplb} enhances the leading order term to  $a_0^0=0.20\pm
0.01$ - the correction goes in the
right direction, but the result is still on the low side as far as the
 present experimental value is concerned.
 To decide about agreement/disagreement between theory and experiment,
 one should i)  evaluate
the scattering lengths in the theoretical framework at order $p^6$, and ii)
determine them more precisely experimentally.
Let me first comment on the {\em theoretical} work.
 \begin{center}
$
\hspace{1cm}\begin{array}{ccc}
\framebox{Theory}&\hspace{0cm}&\framebox{Experiment}
\\&&\\
A_2=\frac{s-M_\pi^2}{F_\pi^2}&& K\rightarrow \pi\pi e \nu
\\
&&\mbox{30 000 decays}\\
\left.
\begin{array}{c} \\ \end{array}
\right\downarrow &&
\left.
\begin{array}{c} \\ \end{array}
\right\downarrow
\\
a_0^0=0.16 && a_0^0=0.26\pm 0.05
\\&&\\
\hspace{-1.5cm}\left.
\begin{array}{c}+\;O(p^4)  \\ \end{array}\right\downarrow&&
\hspace{1.3cm}\left.\hspace{-1.5cm}\begin{array}{c} {\mbox{DIRAC}}
\end{array}\right\Downarrow\begin{array}{c}
{\mbox{E865}}\\{\mbox{KLOE}}\end{array}
\\&&\\
a_0^0=0.20\pm 0.01&&\framebox{$a_0^0$=\mbox{\Large{?}}}
\\&&\\
\hspace{-1cm}\left.
\begin{array}{c}+\;O(p^6) \\ \end{array}\right\Downarrow
\begin{array}{c}  \\\end{array}&&
\\
\\
\framebox{$a_0^0$=\mbox{\Large{?}}}&&
\end{array}
$
\end{center}
Figure 2. Progress in the determination of the elastic $\pi\pi$ scattering
amplitude. References are provided in the text.\\[.5cm]

 The low-energy
expansion of the $\pi\pi$ scattering amplitude is of the form
\bea
A(s,t,u)=
A_2+A_4+A_6 +O(p^8)\co
\eea
where  $A_n$ is of order $p^n$.
The tree-level result $A_2$ is given in (\ref{eqatree}),
 and
the one-loop expression  $A_4$ may be found in \cite{glplb}. The
two-loop contribution $A_6$ was worked out in \cite{pipi6}.
The amplitude $A_2+A_4+A_6$ contains several of the low-energy
 constants from
${\calleff}$:
\bea
\begin{array}{ll}
\left.\begin{array}{l}
 {\cal{L}}_2:F_\pi, M_\pi\\
 {\cal{L}}_4:\bar{l}_1,\bar{l}_2,\bar{l}_3,\bar{l}_4\\
 {\cal{L}}_6: \bar{r}_1,\ldots, \bar{r}_6\end{array}\right\}&
\begin{array}{l}
{\mbox{occur in}}
\;\pi\pi\rightarrow\pi\pi\\
 {\mbox{ at order $p^6$}}.\end{array}
\end{array}
\label{eq2.1}
 \eea

Once the amplitude is available in algebraic form, it is a trivial
matter to  evaluate  the threshold
parameters. To quote an example, the isospin zero
S-wave scattering length is of the form
\bea
a_0^0&=&\frac{7M_\pi^2}{32\pi F_\pi^2}\left\{1+c_4 x +c_6 x^2
\right\} +O(p^8)\, , \nn
x&=&\frac{M_\pi^2}{16\pi^2F_\pi^2}\, .
\eea
The coefficients  $c_4,c_6$ contain
the low-energy constants listed in (\ref{eq2.1}).
Similar formulae hold for all other threshold parameters - the
explicit expressions  for the scattering lengths and effective ranges of the
S-and P-waves as well as for the D-wave scattering lengths at order
$p^6$ may be found
in \cite{pipi6}. It is clear that, before a numerical value for these
parameters can be given, one needs an estimate of the LEC's.
 The calculation is under way - it is, however, quite
involved: One has to solve numerically the Roy-equations \cite{roy} with input
from the high-energy absorptive part. Second, one assumes that the
couplings that describe the mass dependence of the amplitude may be
estimated e.g. from resonance exchange. Requiring that the experimental
amplitude agrees near threshold with the chiral representation allows
one finally to pin down the remaining couplings, as well as the
scattering lengths $a_0^0$ and $a_0^2$. The remaining threshold
parameters may then be obtained from the Wanders sum rules \cite{wanders}.
The first part of the program is completed, and the report will
appear soon \cite{acgl}. The second
part, that will allow us to predict the values of all threshold
parameters, is under investigation \cite{pipi6a}.

On the {\it experimental} side, several  attempts are under way to
 improve our knowledge of the threshold parameters.
 The most
promising ones among them are i) semileptonic $K_{l4}$ decays with improved
 statistics, E865 \cite{e865} and KLOE \cite{kloe},
 and ii) the   measurement of the pionium lifetime -
 DIRAC \cite{dirac} - that will allow one to directly determine
 the combination $|a_0^0-a_0^2|$ of
 S-wave scattering lengths.

Why are we  interested in a precise value of the scattering length
$a_0^0$? First, it is one of the few occasions that
 a quantity in QCD
 can be predicted rather precisely - which is, of course, by itself worth
checking. Second, as has been pointed out in
 \cite{gchpt}, this prediction assumes that the condensate has the
standard size in the chiral limit - in particular, it is assumed to
be non vanishing. For this reason, the authors of Ref.~\cite{gchpt} have
reversed the argument and have set up a framework where the condensate
is allowed to be small or even vanishing in the chiral limit - the
so called generalized chiral perturbation theory.
 [There is
no sign for a small
condensate in present lattice calculations \cite{lattice}.
For further  investigations of the small condensate  scenario see
\cite{derafael,kogan}.]
 Whereas the S-wave
scattering lengths cannot be predicted in this framework, one may
relate their size to the value of  the condensate. Hence, measuring
 $a_0^0$,  $a_0^2$ or a combination thereof \cite{dirac}
 may allow one to determine the
nature of chiral
symmetry breaking by experiment \cite{gchpt,pipigchpt}.

\subsection{Hadronic atoms}

 Using the effective lagrangian framework proposed by Caswell and Lepage some
 time ago \cite{caswell}, the width of pionium in its ground state has
 been determined \cite{gglr} at leading and next-to-leading order in isospin
 breaking and to all orders in the chiral expansion.
  This result will
 allow one to evaluate the combination $|a_0^0-a_0^2|$ with high
 precision, provided that DIRAC determines the lifetime at the 10\%
 level, as is foreseen \cite{dirac}. The technique of Caswell and
 Lepage is very well suited for this purpose, and it is rather easy to
 carry it over to the case of pion nucleon bound states. Work on this
 problem is in progress \cite{pionnucleon}. I refer the reader to
 Soto's contribution to this conference for an outline of the method.

\section*{Acknowledgement}
I wish to thank Stephan Narison and his collaborators for the most
pleasant and interesting conference and for the friendly atmosphere.

\end{document}